\documentclass[aps,prd,a4paper,onecolumn,amsmath,showpacs,superscriptaddress,nofootinbib,preprintnumbers,notitlepage]{revtex4-1}

\usepackage{verbatim}
\usepackage[T1]{fontenc}
\usepackage[utf8]{inputenc}
\usepackage[american]{babel}
\usepackage{epsfig}
\usepackage{graphicx,subcaption,caption}
\usepackage{hyperref}
\usepackage{natbib}
\usepackage{subfiles}
\usepackage{pgfplots}
\usepackage{booktabs}
\usepackage{multirow}
\usepackage{dcolumn}
\usepackage{amsmath}
\usepackage{mathtools}
\usepackage{amsfonts}
\usepackage{amssymb}
\usepackage{epstopdf}
\usepackage{bm}
\usepackage{siunitx}
\usepackage{braket}
\usepackage{enumitem}
\usepackage{soul}
\usepackage{color}
\usepackage{transparent}
\usepackage{pifont}
\usepackage[font={small}]{caption}

\definecolor{navyblue}{rgb}{0.0, 0.0, 0.5}
\definecolor{royalblue}{rgb}{0.25, 0.41, 0.88}
\definecolor{cadmiumgreen}{rgb}{0.0, 0.42, 0.24}
\definecolor{blue-violet}{rgb}{0.54, 0.17, 0.89}
\definecolor{darkviolet}{rgb}{0.58, 0.0, 0.83}
\definecolor{orange(colorwheel)}{rgb}{1.0, 0.5, 0.0}

\usepackage{hyperref}
\hypersetup{
    colorlinks=true, 
    linkcolor=royalblue, 
    citecolor=magenta}

\newcommand\be{\begin{equation}}
\newcommand\ee{\end{equation}}

\newcommand\bea{\begin{eqnarray}}
\newcommand\eea{\end{eqnarray}}

\usepackage{booktabs}
\usepackage{multirow}
\usepackage{dcolumn}
\usepackage{colortbl}

\definecolor{magenta(process)}{rgb}{1.0, 0.0, 0.56}

\definecolor{darkspringgreen}{rgb}{0.09, 0.45, 0.27}

\definecolor{royalblue(web)}{rgb}{0.25, 0.41, 0.88}

\begin{document}

\title{Non-minimal coupling  inflation with constant slow roll}

\author{Mehdi Shokri}
\email{mehdishokriphysics@gmail.com}
\affiliation{Department of Physics, Campus of Bijar, University of Kurdistan, Bijar, Iran}

\author{Jafar Sadeghi}
\email{pouriya@ipm.ir}
\affiliation{Department of Physics, University of Mazandaran, P. O. Box 47416-95447, Babolsar, Iran}

\author{Mohammad Reza Setare}
\email{rezakord@ipm.ir}
\affiliation{Department of Physics, Campus of Bijar, University of Kurdistan, Bijar, Iran}

\author{Salvatore Capozziello}
\email{capozzie@na.infn.it}
\affiliation{Dipartimento di Fisica E. Pancini", Universit\'a di Napoli Federico II", Via Cinthia, I-80126, Napoli, Italy}
\affiliation{Istituto Nazionale di Fisica Nucleare (INFN), sez. di Napoli, Via Cinthia 9, I-80126 Napoli, Italy}
\affiliation{Scuola Superiore Meridionale, Largo S. Marcellino, I-80138, Napoli, Italy}
\affiliation{Laboratory for Theoretical Cosmology, Tomsk State University of
Control Systems and Radioelectronics (TUSUR), 634050 Tomsk, Russia}

\preprint{}
\begin{abstract}
We study a single field inflationary model modified by a non-minimal coupling term between the Ricci scalar $R$ and the scalar field $\varphi$ in the context of constant-roll inflation. The first-order formalism is used to analyse the constant-roll inflation instead of the standard methods used in the literature. In principle, the formalism considers two functions of the scalar field, $W=W(\varphi)$ and $Z=Z(\varphi)$, which lead to the reduction of the equations of motion to first-order differential equations. The approach can be applied to a wide range of cosmological situations since it directly relates the function $W$ with Hubble’s parameter $H$. We perform the inflationary analysis for power-law and exponential couplings, separately. Then, we investigate the features of constant-roll potentials as  inflationary potentials. Finally, we compare the inflationary parameters of the models with the observations of CMB anisotropies in view of realize  a physically motivated model.
\end{abstract}

\maketitle

\section{Introduction}
Inflationary paradigm, as an unavoidable part of modern cosmology, is  the most successful approach to explain  early universe phenomena taking into account the primordial scalar perturbations responsible for large scale structure formation. Also, the primordial gravitational waves are realized as a consequence of tensor perturbations generated during inflationary epoch along with density perturbations \cite{Guth,Kazanas:1980tx,Linde:1981my,Albrecht:1982wi,Lyth:1998xn}. The simplest description of inflation is realized by a single scalar field, the so-called \textit{inflaton}, slowly rolling down  from the peak of a self-interacting  potential to the minimum point in the context of the \textit{slow-roll approximation}. Then, inflation ends up when inflaton decays, at the final phase,  throughout a reheating process \cite{Kofman2,Shtanov}. The observational constraints coming from Planck data have restricted or ruled out a wide range of single field models \cite{martin}. However, some models are still compatible with the observations and nicely reveal the inflationary properties \cite{staro,barrow,kallosh1}. Despite the above interesting features, the single field models suffer from the lack of non-Gaussianity in their spectrum due to uncorrelated modes of the spectrum \cite{Chen}. This could be problematic if observations show  non-Gaussianity in the perturbations spectrum. Consequently, all single scalar field models will be found in an unstable situation. Going beyond the slow-roll approximation, it is proposed to consider some non-Gaussianities in the perturbations spectrum of  single field models. Hence, a new approach to the inflationary paradigm is introduced by considering a scalar field with a constant rate of rolling during the inflationary era, that is 
\begin{equation}
\ddot{\varphi}=\beta H\dot{\varphi}\,,
\label{1}
\end{equation}
where $\beta=-(3+\alpha)$ and $\alpha$ is a non-zero parameter \cite{martin2,Motohashi1,Motohashi2}. Deviation from the slow-roll approximation  can also be found in the realm of \textit{ultra slow-roll} inflation where we deal with a non-negligible $\ddot{\varphi}$ in the Klein-Gordon equation as  $\ddot{\varphi}=3H\dot{\varphi}$. This class of inflationary models shows a finite value for the non-decaying mode of curvature perturbations \cite{Inoue}. Also, the inflationary solutions of the ultra slow-roll model are situated in the non-attractor phase space of inflation but they show a scale-invariant perturbation spectrum. Even, these models sometimes reveal a dynamical  attractor-like behavior \cite{Pattison}. Although the ultra slow-roll model predicts a large $\eta$, it is not able to solve the $\eta$ problem introduced in supergravity for the hybrid inflationary models \cite{Kinney}. Besides, the main problem of ultra models is that the non-Gaussinaity consistency relation of single field models is violated in the presence of ultra condition through Super-Hubble evolution of the scalar perturbation \cite{Namjoo}. As another class of models proposed in the beyond of the slow-roll approximation, we can introduce the \textit{fast-roll} models in which a fast-rolling stage is considered at the beginning of inflation and it is attached to the standard slow-roll only after a few e-folds \cite{Contaldi,Lello,Hazra}. 

Recently, a constant-roll inflationary approach  has been remarkably considered  and one can find a lot  of inflationary models studied in this regime, see, e.g., \cite{Odintsov,Nojiri,Motohashi9,Cicciarella,Anguelova,Ito,karam1,Ghersi,Lin,Micu,Oliveros,Motohashi3,Kamali,diego}. An interesting property of this new viewpoint is that the constant-roll inflation can be unified with some dark energy models described in the context of $f(R)$ gravity \cite{Sebastiani}. Such a unification has also been studied in  other papers \cite{o1,o2,o3}. 

One of most widely-used
inflationary models is the Non-Minimal Coupling (NMC) model in which the Ricci curvature scalar is non-minimally coupled to the scalar field \cite{Spokoiny:1984bd,Lucchin:1985ip, Maeda,Salopek,Fakir,Luca,Kaiser,Shinji,Maria,Kallosh,Edwards,Chiba,Yang,Lotfi,Pieroni,Sami,shokri1,Capozziello:1993xn, Tenkanen:2017jih, Burin2,pi,shokri2}. Hence, the standard inflationary action is modified with a NMC term $\frac{1}{2}\xi
R\varphi^{2}$ as follows
\begin{equation}
S=\int{d^{4}x\sqrt{-g}\bigg(\frac{R}{2}-\frac{1}{2}g^{\mu\nu}\partial_{\mu}\varphi\partial_{\nu}\varphi-V(\varphi)+\frac{1}{2}\xi
R\varphi^{2}\bigg)}\,,
\label{2}
\end{equation}
where $\xi$ is coupling constant and its value is deeply effective on the viability of a given  inflationary model. In principle, NMC arises in  presence of scalar field  quantum corrections.  It is also necessary  for the renormalization of scalar field  in curved space \cite{faraoni11}. Despite the need of using NMC, it can be distractive when we consider a pure mass term $\frac{1}{2}m^{2}\varphi^{2}$ in the action (\ref{2}). In such a case, it is more difficult to obtain the slow-rolling scalar field since the NMC term plays the role of an effective mass for the inflaton.   

The standard attitude in the literature  is that $\xi$ is a free parameter to be fine-tuned in order to solve problems of the inflationary model \cite{faraoni12}. Sometimes this viewpoint is not suitable for our physical situations and we are forced to focus on a more precise approach where $\xi$ is fixed by particle physics prescriptions \cite{faraoni13}. The value of $\xi$, in this attitude, depends on the nature of inflaton and the considered theory of gravity.

Conformal transformations
are adopted to overcome difficulties coming from NMC  since they are often used as a mathematical tool to map the equations of motion  into mathematically equivalent sets of equations that are more easily solved and computationally more convenient to study. Generally, metrics in Jordan and Einstein frames are conformally connected by 
\begin{equation}
\hat{g}_{\mu\nu}=\Omega^{2} g_{\mu\nu}
\label{3}
\end{equation}
where the conformal factor $\Omega=\Omega(\varphi(x))$ is a non-null, differentiable  function. The conformal transformation has been applied for many gravitational theories based on the scalar field since $\Omega$ depends on the scalar field including scalar-tensor and non-linear theories of gravity, Kaluza-Klein theories, fundamental scalar fields e.g. Higgs bosons and dilatons in supergravity theories \cite{faraoni14, Report, Mauro, Bisabr,jarv,Domnech,Catena,Flanagan}. At classical level, seemingly the results of two frames are similar as a consequence of the mathematical equivalence of two frames. However, in the presence of quantum corrections of the scalar field, the physical equivalence between two frames might be broken. In this sense, conformal transformations are not only a mathematical tool but could be related to a deep physical meaning. See, e.g. \cite{Prado, Sergey}.

The  aim of this paper is to study the standard inflationary model equipped with the NMC term in the context of constant-roll inflation. Here we abandon the conventional method used in the recent constant-roll papers and  focus on the first-order formalism to find the suitable potential. The method is  useful to investigate cosmological situations using the obtained scalar field potential \cite{Bazeia1,Bazeia2,Bazeia3,Bazeia4,Bazeia5,Bazeia6,Bazeia7}. This can be achieved by introducing two main functions of the scalar field,  $W=W(\varphi)$ and $Z=Z(\varphi)$,  which lead to reduce the equations of motion to the first-order differential equations. Let us recall  that $W$ is usually called superpotential when we work in the supersymmetry regime. However,   here we do not consider any supersymmetry effects. The above discussion motivates us to arrange the paper as follows. In Section II, we introduce the non-minimal coupling inflationary model in both Jordan and Einstein frames. Section III is devoted to the constant-roll inflationary analysis for two types of coupling, the power-law and the exponential in the first-order formalism. We examine the obtained constant-roll potentials from different points of view to be addressed as the inflationary potentials and also we calculate the inflationary parameters of the models. In section IV, we analyze the obtained results of the models by comparing with the observational datasets coming from the Planck and the BICEP2/Keck array satellites. Conclusions are drawn in Section V.

\section{Non-minimal coupling inflation}
A NMC between  the Ricci scalar $R$ and scalar field $\varphi$ is unavoidable in different cosmological scenarios.  Hence,   the action (\ref{1}) can be generalized as follows \cite{Ritis}
\begin{equation}
S_{J}=\int{d^{4}x\sqrt{-g}\bigg(\frac{f(\varphi)R}{2}-\frac{1}{2}g^{\mu\nu}\partial_{\mu}\varphi\partial_{\nu}\varphi-V(\varphi)\bigg)}
\label{4}
\end{equation}
where $f(\varphi)$ is a generic function of the scalar field and  we assume $\kappa^{2}\equiv8\pi G=1$. By varying the action (\ref{4}) with respect to the metric, the Einstein field equations take the following form
\begin{equation}
f G_{\mu\nu}=T_{\mu\nu}
\label{5}
\end{equation}
with the energy-momentum tensor 
\begin{equation}
T_{\mu\nu}=\nabla_{\mu}\varphi\nabla_{\nu}\varphi-g_{\mu\nu}\bigg(\frac{1}{2}\nabla^{\gamma}\varphi\nabla_{\gamma}\varphi+V(\varphi)\bigg)-\bigg(g_{\mu\nu}\Box f-\nabla_{\mu}\nabla_{\nu}f\bigg)
\label{6}
\end{equation}
where the d'Alembert operator is expressed as ${\displaystyle \Box=\frac{1}{\sqrt{-g}}\partial_{\nu}[\sqrt{-g}g^{\mu\nu}\partial_{\mu}]}$. Let us  rewrite the field equations (\ref{5}) as
\begin{equation}
G_{\mu\nu}=\tilde{T}_{\mu\nu}
\label{7}
\end{equation}
where ${\displaystyle \tilde{T}_{\mu\nu}\equiv\frac{T_{\mu\nu}}{f}}$ and the conservation law of the energy-momentum tensor $\nabla^{\mu}\tilde{T}_{\mu\nu}=0$ is valid as a direct  consequence of the Bianchi identities $\nabla^{\mu}G_{\mu\nu}=0$. It is worth noticing that the present approach introduces some critical values of $\varphi$ extracted from the singularity $f=0$, which are  the barriers that the scalar field cannot cross. Also, by using the energy-momentum tensor of a perfect fluid and Friedman-Robertson-Walker (FRW) metric $ds^{2}=-dt^{2}+a(t)^{2}(dx^{2}+dy^{2}+dz^{2})$ for a  homogeneous, isotropic and spatially flat universe, the dynamical equations are 
\begin{equation}
3fH^{2}=\frac{\dot{\varphi}^{2}}{2}+V-3Hf'\dot{\varphi},
\label{8}
\end{equation}
\begin{equation}
2f\dot{H}=-(1+f'')\dot{\varphi}^{2}-f'(\ddot{\varphi}-H\dot{\varphi})
\label{9}
\end{equation}
where ${\displaystyle H\equiv\frac{\dot{a}}{a}}$ is the Hubble parameter, and the dot and the prime represent the derivative with respect to  cosmic time and $\varphi$, respectively. Moreover, by varying the action (\ref{4}) with respect to  $\varphi$, we find a Klein-Gordon equation 
\begin{equation}
\ddot{\varphi}+3H\dot{\varphi}+\frac{d V}{d\varphi}-3f'(2H^{2}+\dot{H})=0.
\label{10}
\end{equation}
Obviously, the above expressions will be reduced to the standard case as soon as  $f=1$. 
By using the conformal transformation (\ref{3}), we can move from the Jordan to Einstein frame with the conformal factor 
\begin{equation}
\Omega^{2}=f
\label{11}
\end{equation}
and, as a result, the form of action in the Einstein frame takes the  familiar form \cite{Marino}
\begin{equation}
S_{E}=\int
d^{4}x\sqrt{-\hat{g}}\bigg(\frac{\hat{R}}{2}-\frac{1}{2}F^{2}(\varphi)\hat{g}^{\mu\nu}\partial_{\mu}\hat{\varphi}\partial_{\nu}\hat{\varphi}
-\hat{V}(\hat{\varphi})\bigg)
\label{12}
\end{equation}
where $F$ and the potential of the redefined scalar field $\hat{\varphi}$ are expressed as
\begin{equation}
F^{2}(\varphi)\equiv\bigg(\frac{d\hat{\varphi}}{d\varphi}\bigg)^{2}=\frac{2f+3f'^{2}}{2f^{2}},\quad\quad\hat{V}(\hat{\varphi})\equiv\frac{V(\varphi)}{\Omega^{4}}=\frac{V(\varphi)}{f^{2}}.
\label{13}
\end{equation}
The dynamical equations in the Einstein frame take the form 
\begin{equation}
\hat{H}^{2}=\frac{1}{3}\bigg(\frac{1}{2}(\frac{d\hat{\varphi}}{d\hat{t}})^{2}+\hat{V}({\hat{\varphi}})\bigg),\hspace{1cm} \frac{d\hat{H}}{d\hat{t}}=-\frac{1}{2}(\frac{d\hat{\varphi}}{d\hat{t}})^{2},\hspace{1cm} \frac{d^{2}\hat{\varphi}}{d\hat{t}^{2}}+3\hat{H}(\frac{d\hat{\varphi}}{d\hat{t}})
+\frac{d\hat{V}(\varphi)}{d\hat{\varphi}}=0.
\label{14}
\end{equation}
Likewise, the quantities in the two frames are connected by $\hat{a}\equiv\Omega a=\sqrt{f}a$ and $d\hat{t}\equiv\Omega dt=\sqrt{f}dt$. It is worth noticing  that the constant-roll analysis of non-minimal coupling inflation will be performed in the Jordan frame since the constant-roll condition deals with the Hamilton-Jacobi formalism. Then, after finding the constant-roll potential, we can move to the Einstein frame as an easier frame in order to find the observational constraints on the parameters space of the model. 

\section{The First-order formalism}
Let us focus now on the first-order formalism  to analyze  non-minimal constant-roll inflation. The formalism is based on two functions of the scalar field, $W(\varphi)$ and  $Z(\varphi)$, reducing the cosmological equations to  first-order differential equations. It can be applied to a wide range of cosmological situations since it directly relates the function $W$ with Hubble’s parameter $H$. Let us define
\begin{equation}
H=W(\varphi), 
\label{15}
\end{equation}
which allows us to write the relation (\ref{9}) as
\begin{equation}
\dot{\varphi}=\frac{Wf'(1-\beta)-2W'f}{1+f''}
\label{16}
\end{equation}
where prime represents the derivative with respect to  $\varphi$. In such a case, the potential in  expression (\ref{8}) takes the form
\begin{equation}
V=\frac{\mathcal{A}W^{2}+\mathcal{B}WW'-4f^{2}W'^{2}}{2(1+f'')^{2}}
\label{17}
\end{equation}
where
\begin{equation}
\mathcal{A}=6f(1+f'')^{2}+f'^{2}(1-\beta)(5+\beta+6f''),\quad\quad\quad \mathcal{B}=-4ff'(\beta+2+3f'').
\label{18}
\end{equation}
Finally, by using the Klein-Gordon equation (\ref{10}) and  the constant-roll condition (\ref{1}), we obtain the constraint  
\begin{equation}
\mathcal{C}W^{2}+\mathcal{D}WW'+\mathcal{E}W'^{2}+\mathcal{F}WW''+\mathcal{G}W'W''=0
\label{19}
\end{equation}
where
\begin{equation}
\mathcal{C}=f'(1+f'')\bigg(-3(1+f'')^{2}+f''(1-\beta)(5+\beta+6f'')+(1+f'')(3-2\beta-\beta^{2})\bigg)-f'^{2}f'''(1-\beta)\bigg(2+3f''+\beta\bigg),
\label{20}
\end{equation}
\begin{eqnarray}
\mathcal{D}&\!=&\!2f(1+f'')^{2}(3f''-\beta)-2ff''(1+f'')(\beta+2+3f'')\nonumber\\&\!&\!+f'^{2}(1+f'')(-\beta^{2}-3\beta(f''+1)-(3f''+2))+2ff'f'''(1+2\beta+3f'')
\label{21}
\end{eqnarray}
and 
\begin{equation}
\mathcal{E}=-2ff'(1+f'')(\beta+1)+4f^{2}f''',\quad\quad\quad \mathcal{F}=-2ff'(1+f'')(2+\beta+3f''),\quad\quad\quad \mathcal{G}=-4f^{2}(1+f'').
\label{22}
\end{equation}
Now,  the analysis can be specified for two suitable couplings, that is the power-law and the exponential one.

\subsection{The power-law coupling}
First, let us  consider the power-law  as the simplest form of coupling widely-used in  literature as
\begin{equation}
f(\varphi)=1-\xi\varphi^{2} 
\label{23}
\end{equation}
where $\xi$ is a coupling constant and its value  determines the viability of inflationary models. By solving the constraint (\ref{19}), we obtain 
\begin{equation}
W(\varphi)=\mathcal{M}(1-\xi\varphi^{2})^\frac{-\beta-2+6\xi}{2}
\label{24}
\end{equation}
where $\mathcal{M}$ is an integration constant and then the potential (\ref{17}) is given by
\begin{equation}
V(\varphi)=3\mathcal{M}^{2}(1-\xi\varphi^{2})^{(-\beta-2 +6\xi)}\bigg\{1-\frac{\xi\varphi^{2}}{(1-2\xi)^{2}}\bigg(-24\xi^{3}+28\xi^{2}-10\xi+1\bigg)\bigg\}
\label{25}  
\end{equation}
where $\beta=-(3+\alpha)$ and $\alpha$ is a non-zero parameter so that \textit{ultra} and standard \textit{slow-roll} inflation can be driven when $\beta$ is 3 and 0, respectively. Also, the scalar field (\ref{16}) takes the following intrinsic form  
\begin{equation}
-\frac{(1-\xi\varphi^{2})^{\frac{4+\beta-6\xi}{2}}}{4+\beta-6\xi}{}_2 F_1\bigg(1, \frac{(4+\beta-6\xi)}{2}; \frac{(6+\beta-6\xi)}{2}; 1-\xi\varphi^{2}\bigg)=-6\mathcal{M}\xi t
\label{26}    
\end{equation}
where ${}_2 F_1$ is the Hypergeometric function.  
\begin{figure*}[!hbtp]
	\centering
	\includegraphics[width=.355\textwidth,keepaspectratio]{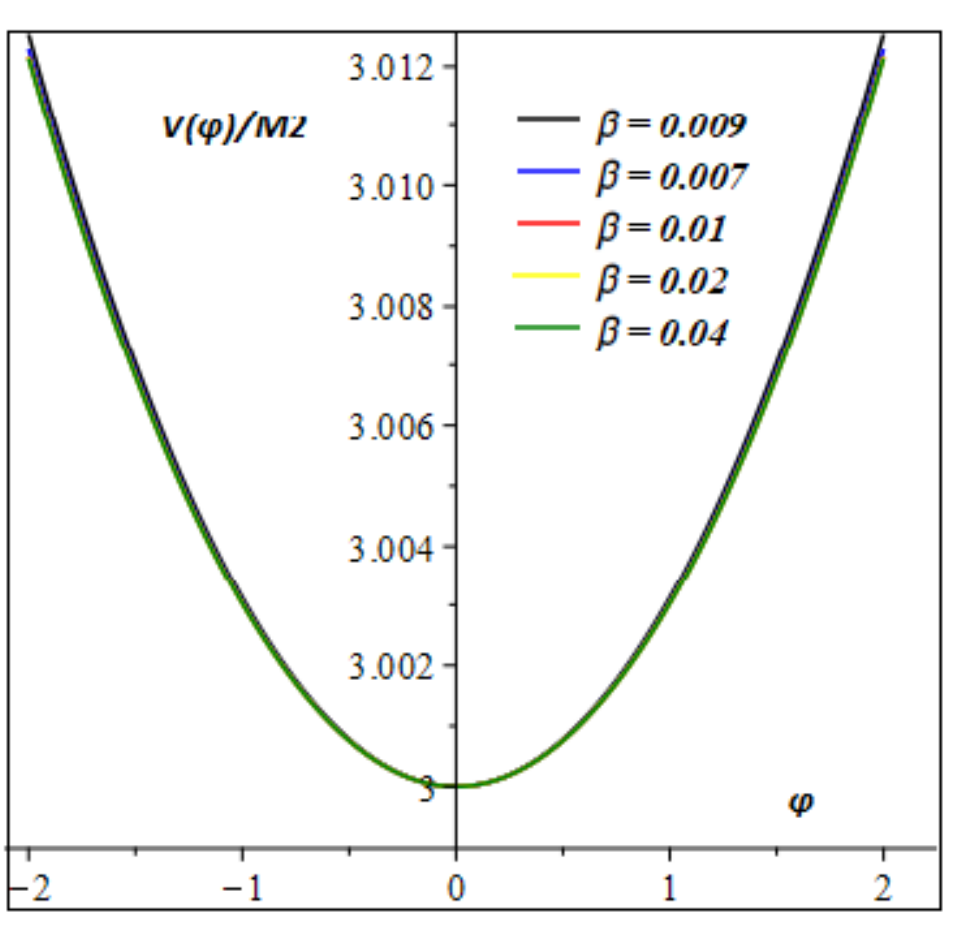}
	\hspace{0.5cm}
	\includegraphics[width=.41\textwidth,keepaspectratio]{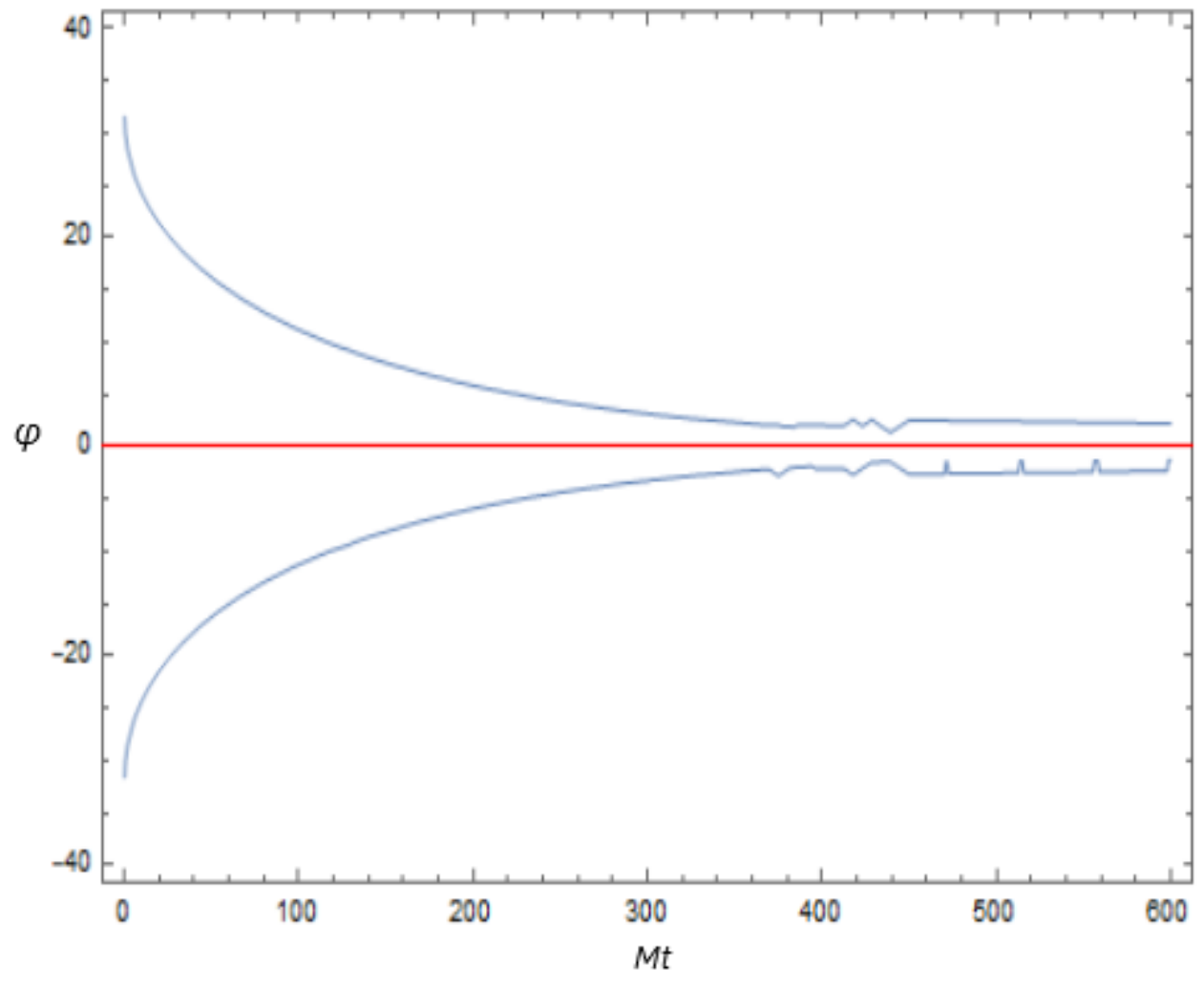}
	\caption{Left panel: The potential (\ref{25}) is plotted for different values of $\beta$ with $\xi=\mathcal{O}(10^{-3})$. Right panel: The evolution of inflaton (\ref{26}) for $\beta=0.01$ and $\xi=\mathcal{O}(10^{-3})$.}
	\label{fig1}
\end{figure*} 

Now, let us continue the analysis by considering  the  plots in Fig.\ref{fig1}. The left panel  presents the potential (\ref{25}) for different values of $\beta$ including $\beta=0.009$, $\beta=0.007$, $\beta=0.01$, $\beta=0.02$ and $\beta=0.04$ when $\xi=\mathcal{O}(10^{-3})$. We can see that the potential behaves like a typical chaotic inflationary potential as the generator of large-field inflationary models for all values of $\beta$. Also, the right panel of Fig. \ref{fig1} yields the evolution of inflaton (\ref{26}) during the inflationary era for $\beta=0.01$ and $\xi=\mathcal{O}(10^{-3})$. Obviously, the inflaton value monotonically evolves to the minimum value and then for large values of $t$, it begins to oscillate in order to decay and create particles through a reheating process.

In order to develope the inflationary analysis, we need to calculate the slow-roll parameters defined in the Hamilton-Jacobi Formalism as
\begin{equation}
\epsilon\equiv2\Big(\frac{H'}{H}\Big)^{2},\quad\quad\quad
\eta\equiv\frac{2H''}{H},\quad\quad\quad
\zeta^{2}\equiv\frac{4H'H'''}{H^{2}}
\label{27}
\end{equation}
where the prime implies  the derivation with respect to the scalar field $\varphi$ and inflation ends when the conditions $\epsilon=1$ or $\eta=1$ are fulfilled. Using  Eq. (\ref{24}), the slow-roll parameters are 
\begin{equation}
\epsilon=\frac{2\xi^{2}\varphi^{2}\big(6\xi-\beta-2\big)^{2}}{(1-\xi\varphi^{2})^{2}},   
\label{28}
\end{equation}
\begin{equation}
\eta=\frac{2\xi\big(6\xi-\beta-2\big)\big(6\xi^{2}\varphi^{2}-(\beta+3)\xi\varphi^{2}-1\big)}{(1-\xi\varphi^{2})^{2}},    
\label{29}
\end{equation}
and
\begin{equation}
\zeta^{2}=\frac{4\xi^{3}\varphi^{2}\big(6\xi-\beta-4\big)\big(6\xi-\beta-2\big)^{2}\big(6\xi^{2}\varphi^{2}-(\beta+3)\xi\varphi^{2}-3\big)}{(1-\xi\varphi^{2})^{4}}.
\label{30}
\end{equation}
Also, the number of e-folds is defined as
\begin{equation}
N=\int^{\varphi_{i}}_{{\varphi}_{f}}{\frac{1}{\sqrt{2\epsilon}}}d{\varphi}
\label{31}
\end{equation}
and  being  ($\varphi_{f}\ll\varphi_{i}$) when the inflation ends, the number of e-folds of the model is found to be  
\begin{equation}
N\simeq\frac{\varphi_{i}^{2}}{4\big(2+\beta-6\xi\big)}.
\label{32}
\end{equation}
Finally, the spectral parameters, \textit{i.e.} the first-order of spectral index, the first order of running spectral index and the tensor-to-scalar ratio are defined by 
\begin{equation}
n_{s}=1-6\epsilon+2\eta,\quad\quad\quad
\alpha_{s}=\frac{dn_{s}}{d\ln k}=16\epsilon\eta-24\epsilon^{2}-2\zeta^{2},\quad\quad\quad r=16\epsilon
\label{33}  
\end{equation}
For the present  model we have
\begin{eqnarray}
&\!&\!n_{s}=\frac{1}{\Big(1-4N\xi(2+\beta-6\xi)\Big)^{2}}\bigg(1+6912N\xi^{5}+576N\big(N-6\beta-11\big)\xi^{4}-192N(\beta+2)\big(-3\beta+N-5\big)\xi^{3}+\nonumber\\&\!&\!
+\Big(-24+16(\beta+2)^{2}N^{2}+\big(-32\beta^{3}-176\beta^{2}-320\beta-144\big)N\Big)\xi^{2}-8(\beta+2)(N-\frac{1}{2})\xi\bigg),
\label{34}    
\end{eqnarray}
\begin{eqnarray}
&\!&\!\alpha_{s}=-\frac{128N\xi^{3}\big(6\xi-\beta-2\big)^{3}}{\Big(1-4N\xi(2+\beta-6\xi)\Big)^{4}}\bigg(1080N\xi^{4}-180N(3\beta+5)\xi^{3}+6N\big(15\beta^{2}+50\beta+42\big)\xi^{2}-\nonumber\\&\!&\!
-5\Big(\frac{3}{2}+N(\beta+2)(5\beta^{2}+15\beta+12)\Big)\xi+\frac{5}{4}\beta+1\bigg),
\label{35}      
\end{eqnarray}
and
\begin{equation}
r=\frac{128N\big(2+\beta-6\xi\big)^{3}\xi^{2}}{\Big(1-4N\xi(2+\beta-6\xi)\Big)^{2}}.
\label{36}      
\end{equation}
Also, the consistency relations of the model can be expressed by
\begin{eqnarray}
&\!&\!n_{s}=\frac{\big(2+\beta-6\xi\big)\xi}{2\Big(\sqrt{\xi(6\xi-\beta-2)^{2}\big(288\xi^{3}-(96\beta+192)\xi^{2}+8\xi(\beta+2)^{2}+r\Big)}+24\xi(6\xi-\beta-2)^{2}\Big)^{2}}\times\nonumber\\&\!&\!
\times\Bigg\{\sqrt{2\xi(6\xi-\beta-2)^{2}\Big(288\xi^{3}-(96\beta+192)\xi^{2}+8\xi(\beta+2)^{2}+r\Big)}\big((12r-48)\xi+(-2r+8)\beta-3r+16\big)+\nonumber\\&\!&\!
+12(144r-576)\xi^{4}+ 12\big((-72r+288)\beta-132r+576\big)\xi^{3}+144\big(3(r-4)\beta+5r-24\big)(\beta+2)\xi^{2}
+\nonumber\\&\!&\!
+4\xi\Big(-2(r-4)\beta^{3}+(-11r+48)\beta^{2}+(-20r+96)\beta+64-15r+\frac{3r^{2}}{4}\Big)-\frac{\big((r-4)\beta+r-8\big)r}{2}\Bigg\}
\label{37}    
\end{eqnarray}
and
\begin{eqnarray}
&\!&\!\alpha_{s}=\frac{\big(2+\beta-6\xi\big)^{2}r^{2}\xi^{2}}{2\Big(\sqrt{2\xi(6\xi-\beta-2)^{2}\big(288\xi^{3}-(96\beta+192)\xi^{2}+8\xi(\beta+2)^{2}+r\Big)}+4\xi(6\xi-\beta-2)^{2}\Big)^{4}}\times\nonumber\\&\!&\!
\times\Bigg\{\sqrt{2\xi(6\xi-\beta-2)^{2}\Big(288\xi^{3}-(96\beta+192)\xi^{2}+8\xi(\beta+2)^{2}+r\Big)}+144\xi^{3}-(48\beta+96)\xi^{2}+4(\beta+2)^{2}\xi+\frac{r}{4}\Bigg\}\times\nonumber\\&\!&\!
\times\Bigg\{\sqrt{2\xi(6\xi-\beta-2)^{2}\Big(288\xi^{3}-(96\beta+192)\xi^{2}+8\xi(\beta+2)^{2}+r\Big)}\big(180\xi^{2}-(60\beta+90)\xi+5\beta^{2}+15\beta+12\big)+\nonumber\\&\!&\!
+25920\xi^{5}-180(168+96\beta)\xi^{4}+36\big(368+420\beta+120\beta^{2}\big)\xi^{3}-12\big(216-\frac{15r}{4}+210\beta^{2}+368\beta+40\beta^{3}\big)\xi^{2}
+\nonumber\\&\!&\!
+4\big(5\beta^{4}+35\beta^{3}+92\beta^{2}+\frac{3}{4}(-5r+144)\beta+48-\frac{15r}{4}\big)\xi+\frac{r}{4}(5\beta^{2}+10\beta+8)\Bigg\}.
\label{38}      
\end{eqnarray}
The previous assumption (\ref{15}), as the simplest relation between $W$ and $H$, sometimes is not sufficient to solve the problems of the model particularly when we deal with quantum corrections. Hence, we can require to consider a new function in addition to $W$ so that the relation (\ref{15}) can be modified as
\begin{equation}
H=W(\varphi)+\gamma\xi Z(\varphi)
\label{39}
\end{equation}
\begin{figure*}[!hbtp]
	\centering		\includegraphics[width=.355\textwidth,keepaspectratio]{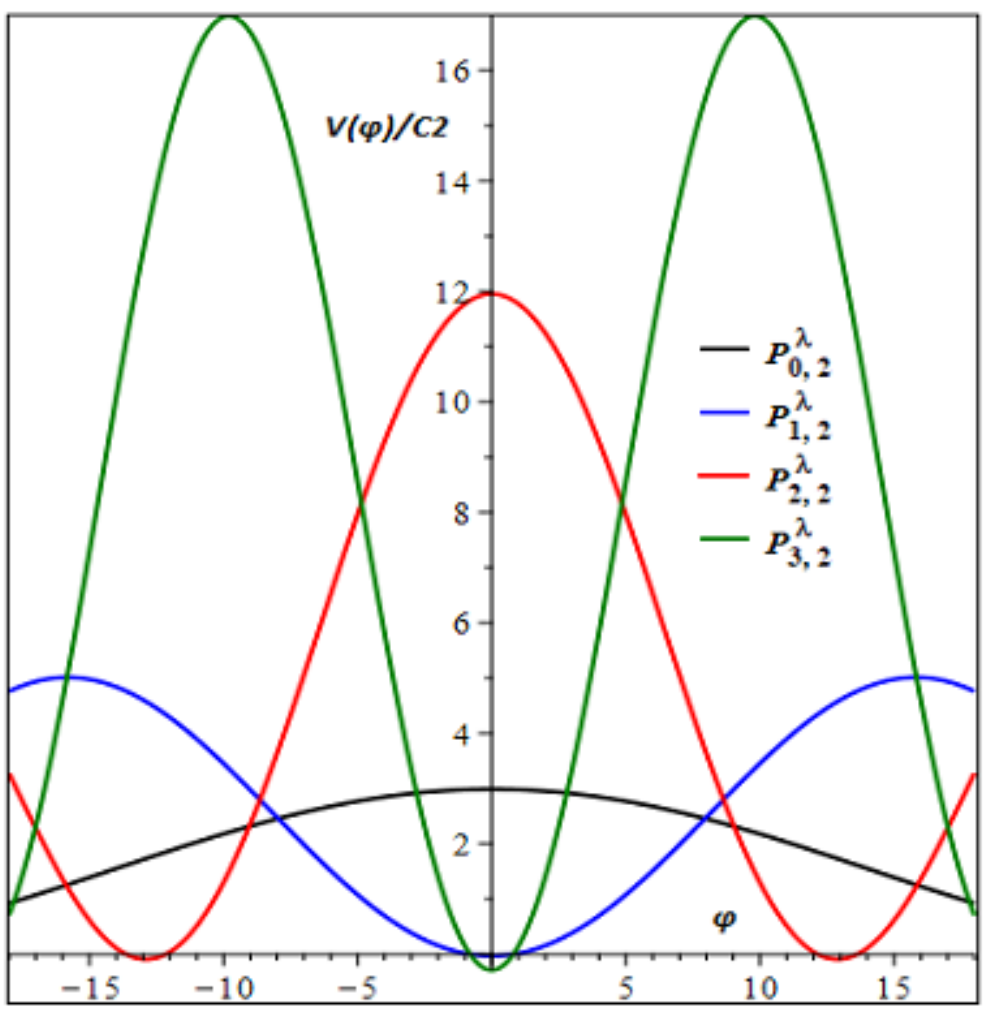}
	\hspace{1.1cm}
	\includegraphics[width=.355
	\textwidth,keepaspectratio]{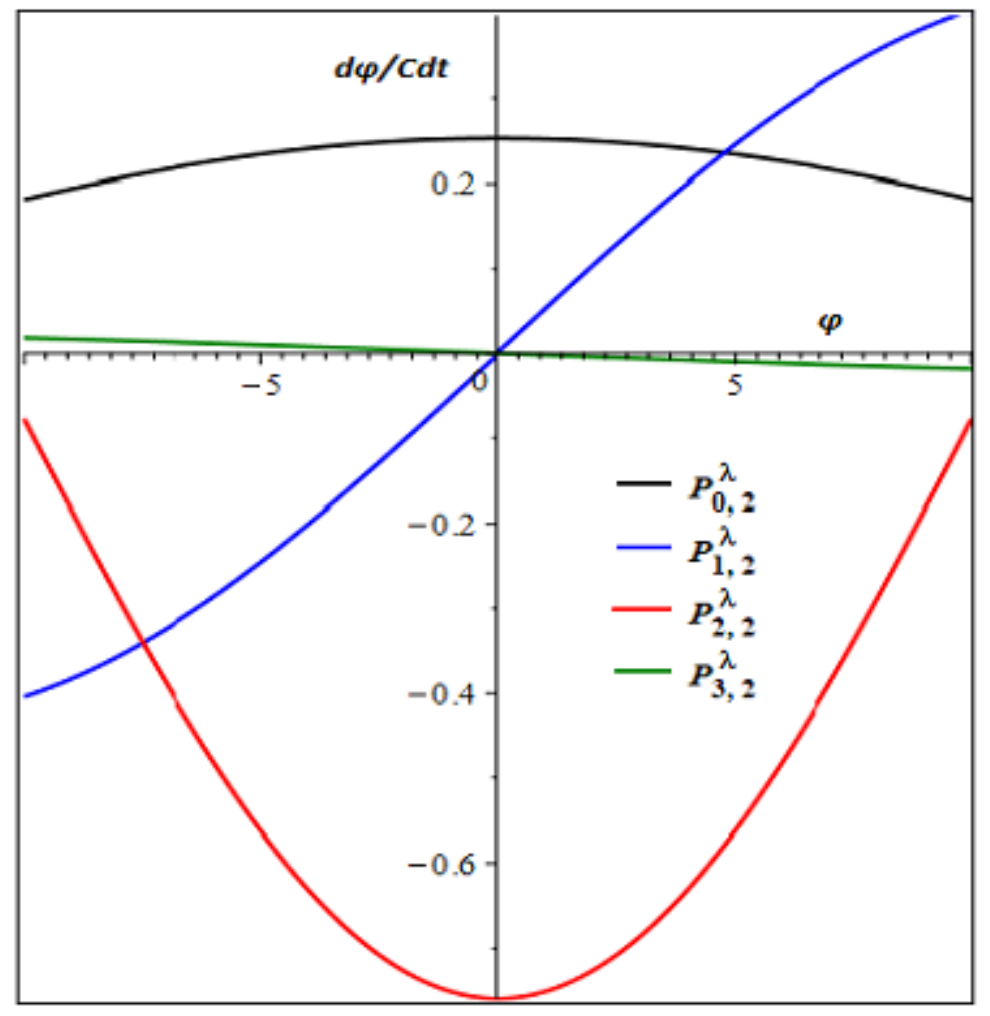}
	\caption{Left panel: The potential (\ref{48}) is plotted for first few Gegenbauer polynomials $P^{\lambda}_{n,2}$ with $\beta=0.01$, $\gamma=1$, $\lambda=2$ and $\xi=\mathcal{O}(10^{-3})$. Right panel: The evolution of inflaton (\ref{50}) for first few Gegenbauer polynomials with  $\beta=0.01$, $\gamma=1$, $\lambda=2$ and $\xi=\mathcal{O}(10^{-3})$.}
	\label{fig2}
\end{figure*}
where $Z=Z(\varphi)$ is an arbitrary function of the scalar field and $\gamma$ is a constant. In such a case, the relation (\ref{16}), the potential (\ref{17}) and the constraint (\ref{19}) are modified as
\begin{equation}
\dot{\varphi}=\frac{(W+\gamma\xi Z)(1-\beta)f'-2(W'+\gamma\xi Z')f}{1+f''},
\label{40}
\end{equation}
\begin{equation}
V(\varphi)=\frac{\mathcal{A}(W+\gamma\xi Z)^{2}+\mathcal{B}(W+\gamma\xi Z)(W'+\gamma\xi Z')-4f^{2}(W'+\gamma\xi Z')^{2}}{2(1+f'')^{2}},
\label{41}
\end{equation}
and
\begin{eqnarray}
&\!&\!\mathcal{C}(W+\gamma\xi Z)^{2}+\mathcal{D}(W+\gamma\xi Z)(W'+\gamma\xi Z')+\mathcal{E}(W'+\gamma\xi Z')^{2}\nonumber\\&\!+&\!
\mathcal{F}(W+\gamma\xi Z)(W''+\gamma\xi Z'')+\mathcal{G}(W'+\gamma\xi Z')(W''+\gamma\xi Z'')=0.
\label{42}
\end{eqnarray}
Let us consider some examples of $Z$. First, we consider $Z=1$ as a trivial case in which the solution of the standard case is reproduced with a tiny modification term as $W=W-\gamma\xi$. Also, the case of $Z=W$ is trivial since it leads to similar solutions with the standard case. As another possibility, we consider $Z=W'$ in which the constraint (\ref{42}) leads to a general differential equation
\begin{equation}
\gamma\xi(1-\xi\varphi^{2})W''+(-\xi\varphi^{2}+\gamma\xi^{2}\varphi(-\beta-2+6\xi)+1)W'+\varphi\xi(-\beta-2+6\xi)W=0. 
\label{43}
\end{equation}
By using the generalized Gegenbauer polynomials  $P^{\lambda}_{n,m}(x)$ 
\begin{equation}
(1-x^{2})P''^{\lambda}_{n,m}(x)-2(\lambda+1)xP'^{\lambda}_{n,m}(x)+\bigg(n(2\lambda+n+1)-\frac{m(2\lambda+m)}{1-x^{2}}\bigg)P^{\lambda}_{n,m}(x)=0  
\label{44}
\end{equation}
in the case of $\xi\varphi^{2}=x^{2}$, we can solve the differential Eq. (\ref{43}) as
\begin{equation}
W(x)=CP^{\lambda}_{n,m}(x)U(x)
\label{45}
\end{equation}
where 
\begin{equation}
P^{\lambda}_{n,m}(x)=\frac{a_{n,m}(\lambda)}{(1-x^{2})^{\lambda+\frac{m}{2}}}(\frac{d}{dx})^{n-m}(1-x^{2})^{\lambda+n},\quad\quad U(x)=(x^{2}-1)^{\frac{-\beta-2+6\xi+2(\lambda+1)}{4}}.
\label{46}
\end{equation}
It is worth noticing that $\lambda > -\frac{1}{2}$ is a given real parameter and $a_{n,m}(\lambda)$ is the normalization coefficient of the generalized Gegenbauer function $P^{\lambda}_{n,m}(x)$. Also, $C$ in (\ref{45}) is the normalization factor, which is obtained by the orthogonal condition. For the rest of the analysis, we consider the case of $m=2$ in which the associated Gegenbauer polynomials $P^{\lambda}_{n,m}(x)$ is connected to the Gegenbauer polynomials $G^{\lambda}_{n}(x)$ and the Hypergeometric functions by
\begin{equation}
P^{\lambda}_{n,2}(x)=G^{\lambda}_{n}(x)=\frac{(2\lambda)_{n}}{n!}{}_2 F_1\bigg(-n,2\lambda+n; \lambda+\frac{1}{2}; \frac{1-x}{2}\bigg).
\label{47}
\end{equation}
Now, the potential (\ref{41}) for $Z=W'$ is
\begin{eqnarray}
&\!&\!V(\varphi)=\frac{C^{2}}{2(1-2\xi)^{2}}\bigg\{\mathcal{A}(P^{\lambda}_{n,2}U)^{2}+\bigg(\gamma^{2}\xi^{2}\mathcal{A}+\gamma\xi\mathcal{B}-4(1-\xi\varphi^{2})^{2}\bigg)(P'^{\lambda}_{n,2}U+P^{\lambda}_{n,2}U')^{2}\nonumber\\&\!+&\!
\bigg(2\gamma\xi\mathcal{A}+\mathcal{B}\bigg)P^{\lambda}_{n,2}U(P'^{\lambda}_{n,2}U+P^{\lambda}_{n,2}U')+\bigg(\gamma^{2}\xi^{2}\mathcal{B}-8\gamma\xi(1-\xi\varphi^{2})^{2}\bigg)(P'^{\lambda}_{n,2}U+P^{\lambda}_{n,2}U')(P''^{\lambda}_{n,2}U+2P'^{\lambda}_{n,2}U'+P^{\lambda}_{n,2}U'')\nonumber\\&\!-&\!
4\gamma^{2}\xi^{2}(1-\xi\varphi^{2})^{2}(P''^{\lambda}_{n,2}U+2P'^{\lambda}_{n,2}U'+P^{\lambda}_{n,2}U'')^{2}+\gamma\xi\mathcal{B}P^{\lambda}_{n,2}U(P''^{\lambda}_{n,2}U+2P'^{\lambda}_{n,2}U'+P^{\lambda}_{n,2}U'')\bigg\}
\label{48}
\end{eqnarray}
where the prime, as above, represents the derivative with respect to the scalar field $\varphi$ and from (\ref{18}), we have
\begin{equation}
\mathcal{A}=2\xi\varphi^{2}\bigg(-12\xi^{2}(3-2\beta)+2\xi(11-4\beta-\beta^{2})-3\bigg)+6(1-2\xi)^{2},\quad\quad \mathcal{B}=8\xi\varphi(\xi\varphi^{2}-1)(6\xi-2-\beta).
\label{49}
\end{equation}
From (\ref{40}), we find
\begin{equation}
\dot{\varphi}=-\frac{C}{(1-2\xi)}\bigg(2\xi\varphi(1-\beta)P^{\lambda}_{n,2}U+(2\gamma\xi^{2}\varphi(1-\beta)+2(1-\xi\varphi^{2}))(P'^{\lambda}_{n,2}U+P^{\lambda}_{n,2}U')+2\gamma\xi(1-\xi\varphi^{2})(P''^{\lambda}_{n,2}U+2P'^{\lambda}_{n,2}U'+P^{\lambda}_{n,2}U'')\bigg).
\label{50}    
\end{equation}
In the left panel of Fig. \ref{fig2}, we show the behavior of the potential (\ref{48}) for first few Gegenbauer polynomials $P^{\lambda}_{0,2}$, $P^{\lambda}_{1,2}$, $P^{\lambda}_{2,2}$, and $P^{\lambda}_{3,2}$ when $\beta=0.01$, $\gamma=1$, $\lambda=2$ and $\xi=\mathcal{O}(10^{-3})$. The potential for the cases of $1,2$ and $3,2$ behaves like a large field inflationary potential in contrast with the cases of $0,2$ and $2,2$ which show a small-field behavior. Also, the plot predicts that by moving to the higher Gegenbauer polynomials, the potential will be more localized around the origin. The right panel of Fig. \ref{fig2} presents the evolution of inflaton (\ref{50}) for the mentioned cases of the Gegenbauer polynomials when $\beta=0.01$, $\gamma=1$, $\lambda=2$ and $\xi=\mathcal{O}(10^{-3})$.
\subsection{The exponential coupling}
Another interesting case  is the constant-roll non-minimal inflation with the exponential coupling between $R$ and $\varphi$ defined by
\begin{equation}
f(\varphi)=e^{k\varphi}   
\label{51}
\end{equation}
where $k$ is a constant. From the constraint (\ref{19}), we obtain
\begin{equation}
ln W(\varphi)=\frac{-3k^{2}e^{k\varphi}-(\beta+2)k\varphi}{2},
\label{52}
\end{equation}
which by using the exponential expansion, $W$ takes the following form 
\begin{equation}
W(\varphi)=\mathcal{M}\exp{\bigg(\frac{-2k\varphi(3k^{2}+\beta+2)-6k^{2}-3k^{4}\varphi^{2}}{4}\bigg)}\,
\label{53}
\end{equation}
where $\mathcal{M}$ is an integration constant. Now, we obtain the potential (\ref{17}) as
\begin{eqnarray}
&\!&\!V(\varphi)=\frac{\mathcal{M}^{2}}{2(1+k^{2}e^{k\varphi})^{2}}\bigg\{6k^{4}(3k^{3}\varphi+3k^{2}+4)\exp\bigg({\frac{-6k^{2}-2k\varphi(3k^{2}+\beta-1)-3k^{4}\varphi^{2}}{2}}\bigg)\nonumber\\&\!&\!-3k^{2}(3k^{6}\varphi^{2}+6k^{5}\varphi+3k^{4}-7)\exp\bigg({\frac{-6k^{2}-2k\varphi(3k^{2}+\beta)-3k^{4}\varphi^{2}}{2}}\bigg)\bigg\}.
\label{54}
\end{eqnarray}
The intrinsic expression for the scalar field (\ref{16}) is 
\begin{equation}
\sqrt{\frac{\pi}{3}}\exp\bigg({\frac{18k^{4}-(3k^{2}+\beta)^{2}}{12k^{2}}}\bigg)\mbox{erfi}\bigg(\frac{3k^{3}\varphi+3k^{2}+\beta}{2\sqrt{3}k}\bigg)=3\mathcal{M}k^{3}t
\label{55}   
\end{equation}
where $\mbox{erfi}$ is the imaginary error function.
\begin{figure*}[!hbtp]
	\centering
	\includegraphics[width=.355\textwidth,keepaspectratio]{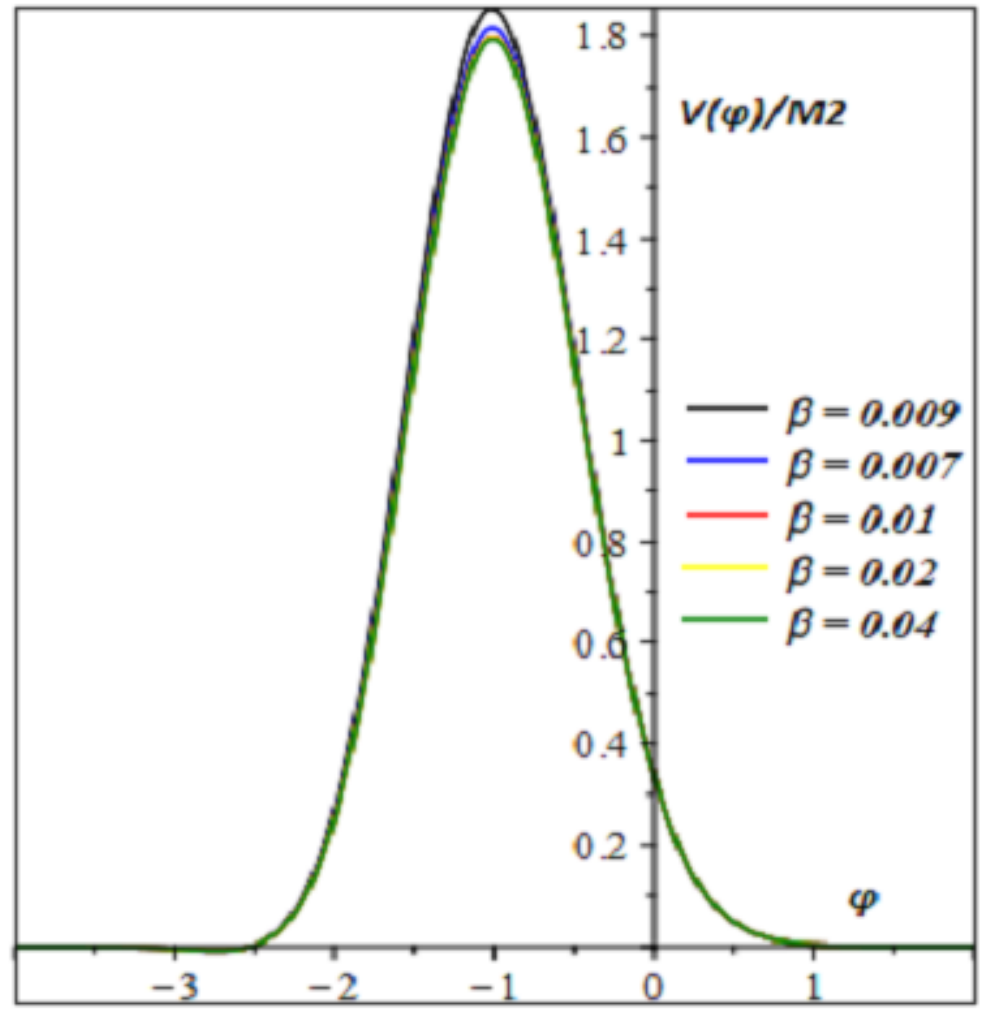}
	\hspace{0.5cm}\includegraphics[width=.378
	\textwidth,keepaspectratio]{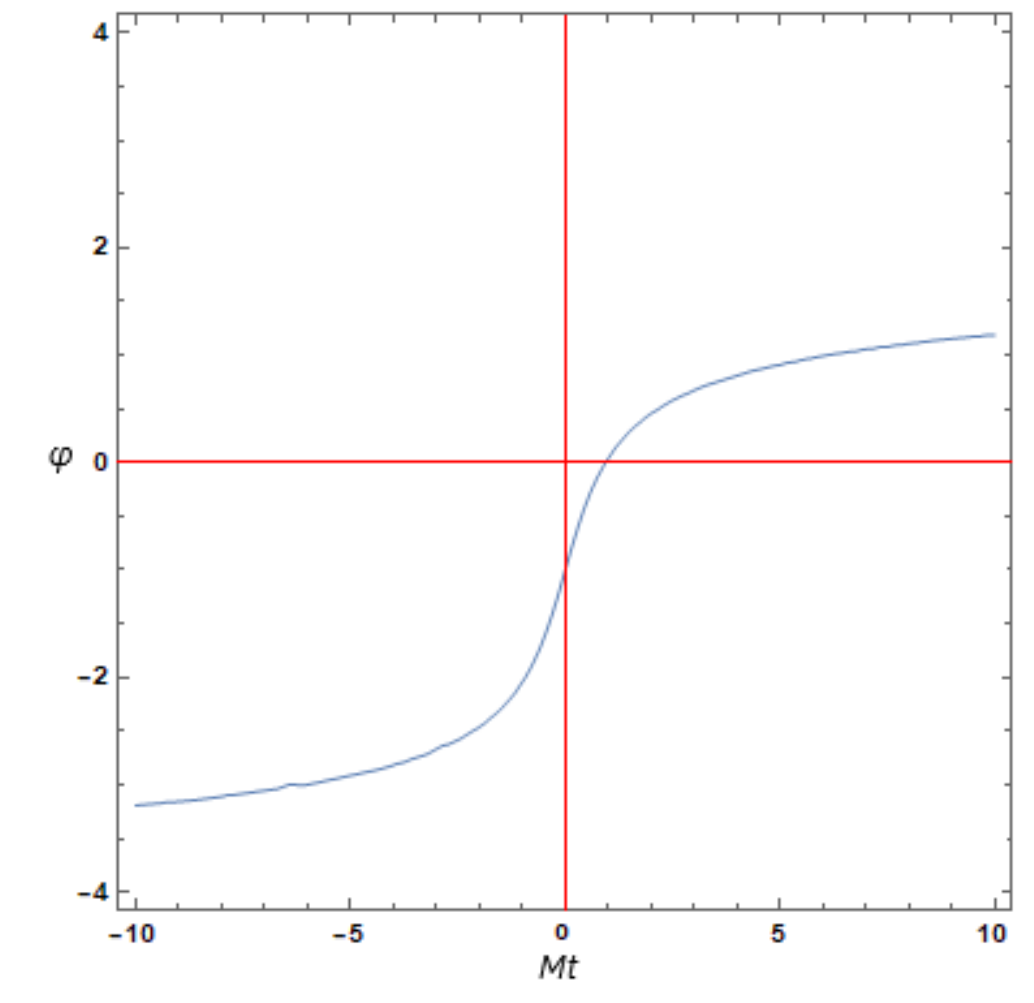}
	\caption{Left panel: The potential (\ref{54}) is plotted for different values of $\beta$ with $k=1$. Right panel: The evolution of inflaton (\ref{55}) for $\beta=0.01$ and $k=1$.}
	\label{fig3}
\end{figure*} 

Considering the obtained plots, the left panel of Fig. \ref{fig3} represents the potential (\ref{54}) for different values of $\beta$ including $\beta=0.009$, $\beta=0.007$, $\beta=0.01$ , $\beta=0.02$ and $\beta=0.04$ when $k=1$. The potential has a Gaussian behavior around the point of $-1$ not the origin. Moreover, by focusing on the non-imaginary part of the right panel of Fig. \ref{fig3}, we find that the value of inflaton increases to the maximum value in the last stages of inflation. Consequently, the potential (\ref{54}) behaves like a small-field inflationary potential.

Now, let us focus on the inflationary parameters of the model. Using  Eq. (\ref{53}), the slow-roll parameters (\ref{27}) are obtained as
\begin{equation}
\epsilon=\frac{1}{2}\Big(3k^{4}\varphi+k(3k^{2}+\beta+2)\Big)^{2},
\label{56}    
\end{equation}
\begin{equation}
\eta=\frac{k^{2}}{2}\Big(9k^{6}\varphi^{2}+18k^{5}\varphi+9k^{4}+6k^{3}\varphi(\beta+2)+6k^{2}(\beta+1)+(\beta+2)^{2}\Big),
\label{57}    
\end{equation}
and
\begin{equation}
\zeta^{2}=\frac{k^{4}}{4}\big(3k^{3}\varphi+3k^{2}+(\beta+2)\big)^{2}\Big(9k^{6}\varphi^{2}+18k^{5}\varphi+9k^{4}+6k^{3}\varphi(\beta+2)+6k^{2}(\beta-1)+(\beta+2)^{2}\Big)
\label{58}    
\end{equation}
Also, the number of e-folds (\ref{31}) is calculated as 
\begin{equation}
N\simeq\frac{1}{3k^{5}(3k^{2}+\beta+2)}\ln\Big({3k^{4}\varphi_{i}+k(3k^{2}+\beta+2)}\Big).
\label{59}
\end{equation}
In this case, the spectral parameters (\ref{33}) are given by
\begin{equation}
n_{s}=1-6k^{4}-2e^{6Nk^{5}(3k^{2}+\beta+2)},\quad\quad\alpha_{s}=-\frac{5e^{12Nk^{5}(3k^{2}+\beta+2)}}{2}-15k^{4}e^{6Nk^{5}(3k^{2}+\beta+2)},\quad\quad r=8e^{6Nk^{5}(3k^{2}+\beta+2)}
\label{60}    
\end{equation}
and the consistency relations take the following forms
\begin{equation}
n_{s}=1-\frac{r}{4}-6k^{4},\quad\quad\alpha_{s}=-\frac{5}{128}r^{2}-\frac{15}{8}k^{4}r.
\label{61}    
\end{equation}
Similar to the power-law coupling case, we can examine  the presence of a new function $Z$. It is   
\begin{equation}
H=W(\varphi)+\gamma kZ(\varphi)
\label{62}
\end{equation}
where again $\gamma$ is a constant. Also, the relation (\ref{16}), the potential of scalar field (\ref{17}) and the constraint (\ref{19}) are expressed as
\begin{equation}
\dot{\varphi}=\frac{(W+\gamma kZ)f'(1-\beta)-2(W'+\gamma kZ')f}{1+f''},
\label{63}
\end{equation}
\begin{figure*}[!hbtp]
	\centering		\includegraphics[width=.355\textwidth,keepaspectratio]{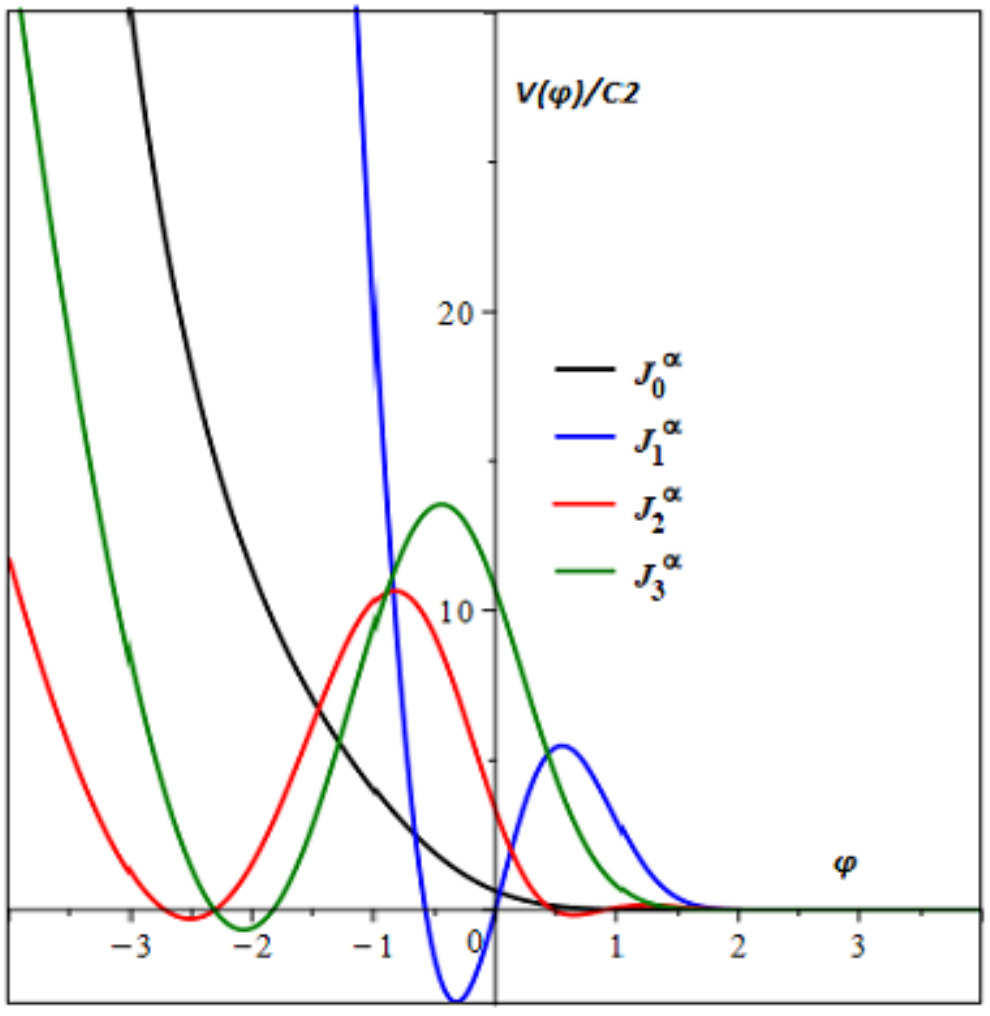}
	\hspace{1.1cm}\includegraphics[width=.355
	\textwidth,keepaspectratio]{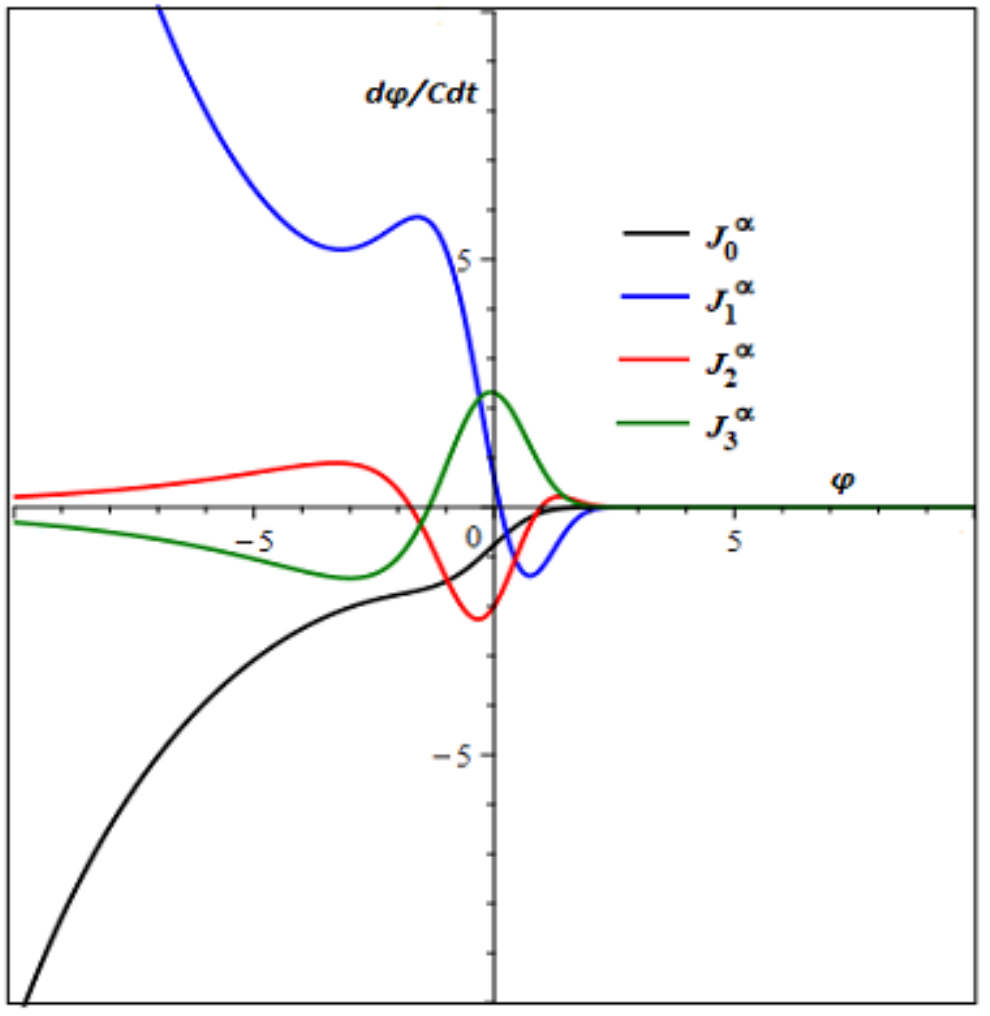}
	\caption{Left panel: The potential (\ref{70}) is plotted for first few polynomials $J^{\alpha}_{n}$ with $\beta=0.01$, $\gamma=k=1$, $\alpha=-4$. Right panel: The evolution of inflaton (\ref{72}) for first few Gegenbauer polynomials with $J^{\alpha}_{n}$ with $\beta=0.01$, $\gamma=k=1$, $\alpha=-4$.}
	\label{fig4}
\end{figure*}
\begin{equation}
V=\frac{\mathcal{A}(W+\gamma kZ)^{2}+\mathcal{B}(W+\gamma kZ)(W'+\gamma kZ')-4f^{2}(W'+\gamma kZ')^{2}}{2(1+f'')^{2}}
\label{64}
\end{equation}
and
\begin{eqnarray}
&\!&\!\mathcal{C}(W+\gamma kZ)^{2}+\mathcal{D}(W+\gamma kZ)(W'+\gamma kZ')+\mathcal{E}(W'+\gamma kZ')^{2}\nonumber\\&\!+&\!
\mathcal{F}(W+\gamma kZ)(W''+\gamma kZ'')+\mathcal{G}(W'+\gamma kZ')(W''+\gamma kZ'')=0.
\label{65}
\end{eqnarray}
The case of $Z=1$ is trivial since the solutions of $H=W$ are slightly corrected as $W=W-\gamma k$. The case of $Z=W$ is also trivial because it  reproduces the results  of $H=W$. In the case $Z=W'$, the constraint (\ref{65}) is solved by
\begin{equation}
2\gamma k(1+k^{2}e^{k\varphi})W''+(3k^{6}\gamma e^{2k\varphi}+k^{2}e^{k\varphi}(2+k^{2}\gamma(\beta+5))+k^{2}\gamma(\beta+2)+2)W'+(3k^{5}e^{2k\varphi}+k^{3}e^{k\varphi}(\beta+5)+k(\beta+2))W=0.
\label{66}
\end{equation}
By using the definition of the polynomials $J^{\alpha,\chi}_{n}(x)$ discussed in \cite{fakhri}
\begin{equation}
(1+x^{2})J''^{\alpha,\chi}_{n}(x)+(\chi+2(\alpha+1)x)J'^{\alpha,\chi}_{n}(x)-(n(n+2\alpha+1))J^{\alpha,\chi}_{n}(x)=0  
\label{67}
\end{equation}
in the case of $k^{2}e^{k\varphi}=x^{2}$, Eq. (\ref{66}) can be solved as
\begin{equation}
W(x)=CJ^{\alpha,\chi}_{n}(x)U(x)
\label{68}
\end{equation}
where $C$ is the normalization factor and 
\begin{equation}
J^{\alpha,\chi}_{n}(x)=\frac{e^{-\chi \arctan(x)}}{n!(1+x^{2})^{\alpha}}(\frac{d}{dx})^n(1+x^{2})^{\alpha+n}e^{\chi\arctan(x)},\quad\quad U(x)=x^{\frac{-(\beta+3)k^{2}\gamma-2}{2\gamma k^{2}}}(x^{2}+1)^{\frac{\alpha+1}{2}}e^{-\frac{3x^{2}}{4}}.
\label{69}
\end{equation}
Note that $\alpha<-1$ and  $-\infty<\chi<\infty$. We restrict ourselves to the polynomials with $\chi=0$ since we want to investigate the issue in the non-imaginary part.
  \begin{figure*}[!hbtp]
	\centering
	\includegraphics[width=.48\textwidth,keepaspectratio]{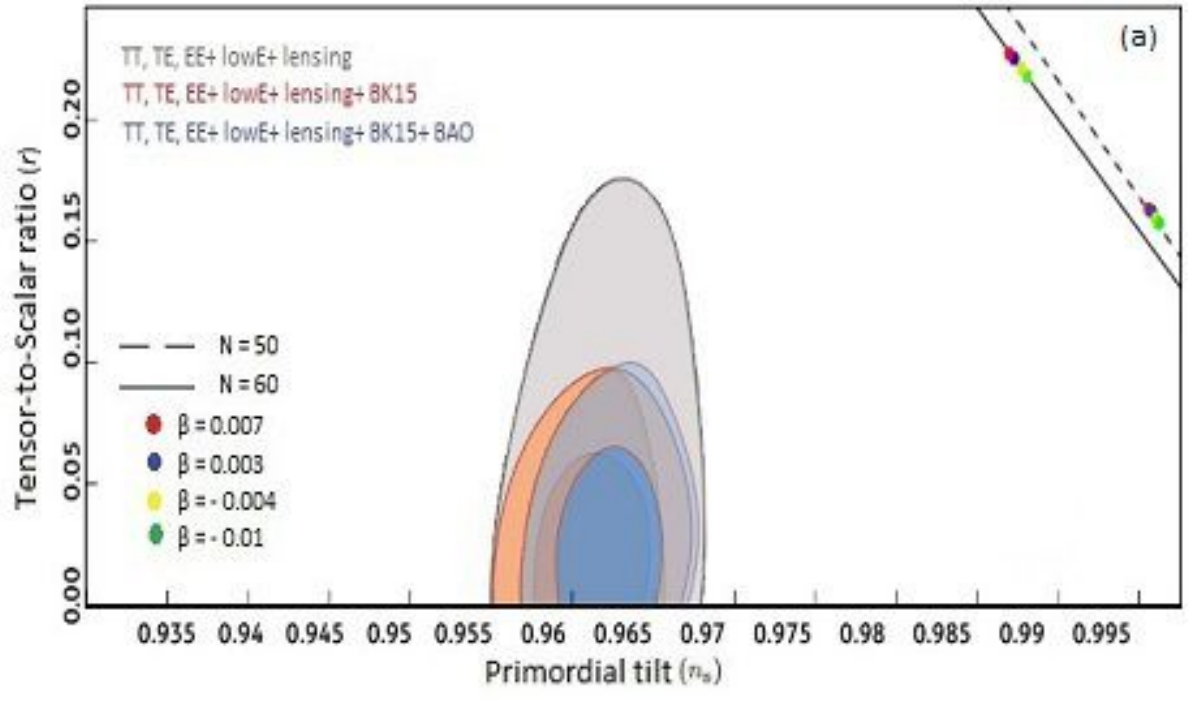}
	\includegraphics[width=.48\textwidth,keepaspectratio]{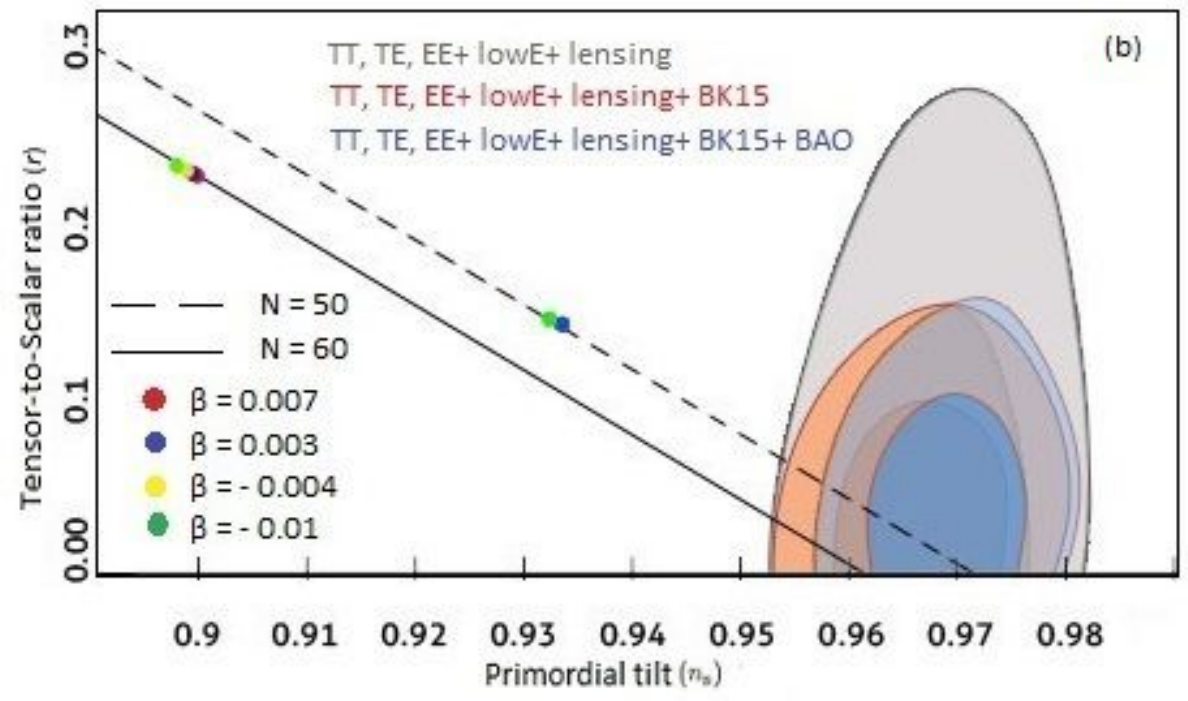}
	\centering		\includegraphics[width=.48\textwidth,keepaspectratio]{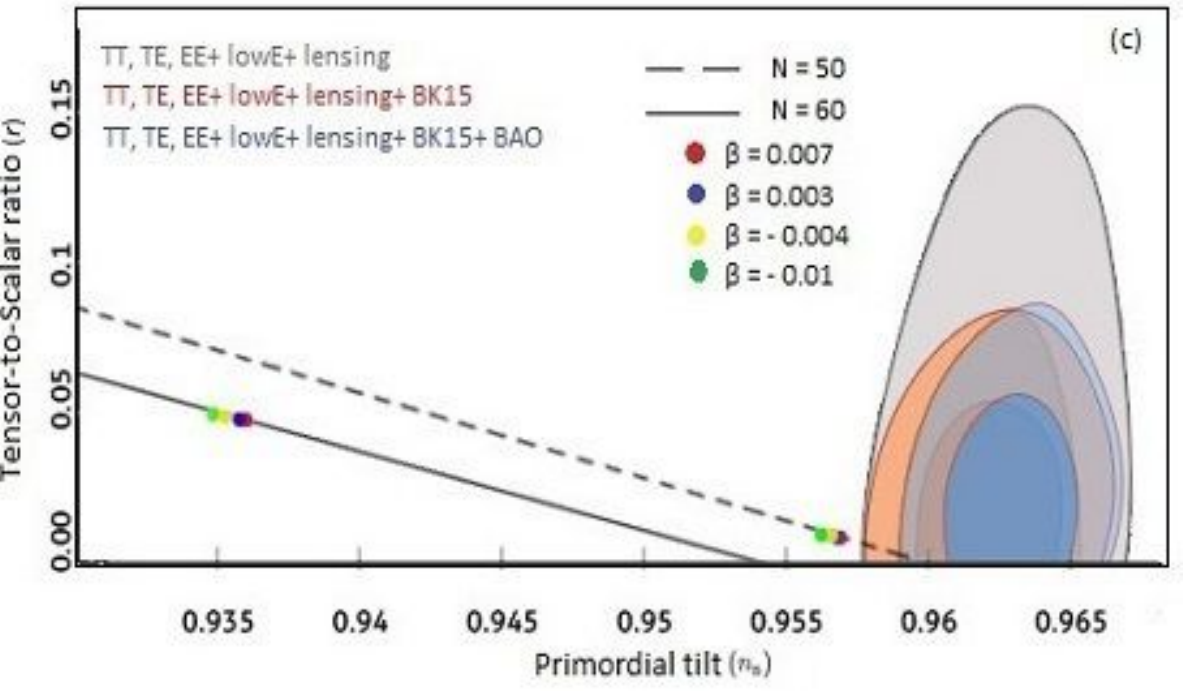}
	\includegraphics[width=.48\textwidth,keepaspectratio]{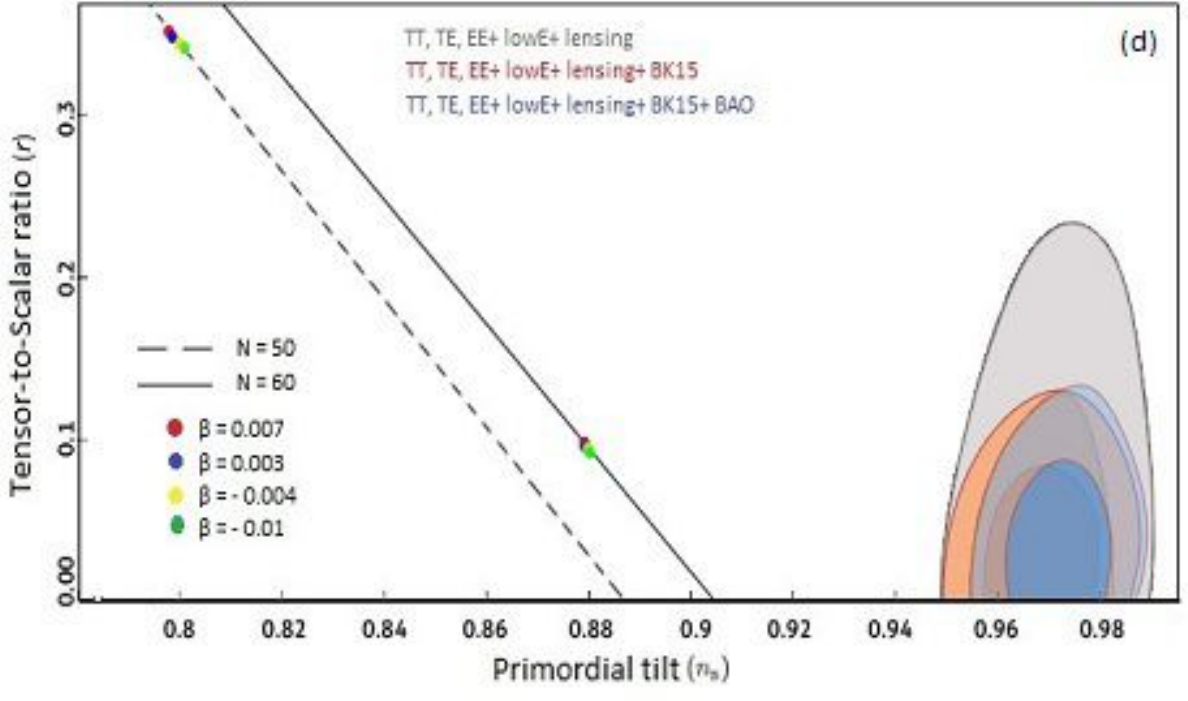}
	\caption{The marginalized joint 68\% and 95\% CL regions for $n_{s}$ and $r$ at $k = 0.002$ Mpc$^{-1}$ from Planck alone and in combination with BK15 or BK15+BAO data \cite{cmb} and the $n_{s}-r$ constraints on the model with a power-law coupling (\ref{23}). The dashed and solid lines represent $N=50$ and $N=60$, respectively. The panel (a) denotes the obtained results of the power-law coupling model with the assumption (\ref{15}) for negative and positive values of $\beta$ with $\xi=\mathcal{O}(10^{-3})$. The panels (b), (c) and (d) denote the obtained results of three first Gegenbauer polynomials $P^{\lambda}_{0,2}$, $P^{\lambda}_{1,2}$ and $P^{\lambda}_{2,2}$ respectively for the power-law model with the assumption (\ref{39}) for negative and positive values of $\beta$ with $\gamma=1$, $\lambda=2$ and $\xi=\mathcal{O}(10^{-3})$.}
	\label{fig5}
\end{figure*}
Now, the potential (\ref{64}) for $Z=W'$ is given by
\begin{eqnarray}
&\!&\!V(\varphi)=\frac{C^{2}}{2(1+k^{2}e^{k\varphi})^{2}}\bigg\{\mathcal{A}(J^{\alpha}_{n}U)^{2}+\bigg(\gamma^{2}k^{2}\mathcal{A}+\gamma k\mathcal{B}-4e^{2k\varphi}\bigg)(J'^{\alpha}_{n}U+J^{\alpha}_{n}U')^{2}\nonumber\\&\!+&\!
\bigg(2\gamma k\mathcal{A}+\mathcal{B}\bigg)J^{\alpha}_{n}U(J'^{\alpha}_{n}U+J^{\alpha}_{n}U')+\bigg(\gamma^{2}k^{2}\mathcal{B}-8\gamma ke^{2k\varphi}\bigg)(J'^{\alpha}_{n}U+J^{\alpha}_{n}U')(J''^{\alpha}_{n}U+2J'^{\alpha}_{n}U'+J^{\alpha}_{n}U'')\nonumber\\&\!-&\!
4\gamma^{2}k^{2}e^{2k\varphi}(J''^{\alpha}_{n}U+2J'^{\alpha}_{n}U'+J^{\alpha}_{n}U'')^{2}+\gamma k\mathcal{B}J^{\alpha}_{n}U(J''^{\alpha}_{n}U+2J'^{\alpha}_{n}U'+J^{\alpha}_{n}U'')\bigg\}
\label{70}
\end{eqnarray}
where, as above,  the prime represents the derivative with respect to the scalar field $\varphi$ and, from Eq. (\ref{18}), it is 
\begin{equation}
\mathcal{A}=6k^{4}e^{3k\varphi}(-\beta+2)+k^{2}e^{2k\varphi}(17-4\beta-\beta^{2})+6e^{k\varphi},\quad\quad \mathcal{B}=-12k^{3}e^{3k\varphi}-4ke^{2k\varphi}(\beta+2).
\label{71}
\end{equation}
Also, from Eq. (\ref{63}), we find
\begin{equation}
\dot{\varphi}=\frac{C}{(1+k^{2}e^{k\varphi})}\bigg(k(1-\beta)e^{k\varphi}J^{\alpha}_{n}U+(\gamma k^{2}(1-\beta)-2)e^{k\varphi}(J'^{\alpha}_{n}U+J^{\alpha}_{n}U')-2\gamma ke^{k\varphi}(J''^{\alpha}_{n}U+2J'^{\alpha}_{n}U'+J^{\alpha}_{n}U'')\bigg).
\label{72}    
\end{equation}
The left panel of Fig. \ref{fig4} shows the behaviour of potential (\ref{70}) for the  first few polynomials $J^{\alpha}_{0}$, $J^{\alpha}_{1}$, $J^{\alpha}_{2}$,
$J^{\alpha}_{3}$ when $\beta=0.01$, $\gamma=k=1$, $\alpha=-4$. Seemingly, the potential in two cases $J^{\alpha}_{2}$,
$J^{\alpha}_{3}$ shows a similar behavior. The right panel of Fig. \ref{fig4} presents the evolution of inflaton (\ref{72}) for the mentioned case of the polynomials when $\beta=0.01$, $\gamma=k=1$, $\alpha=-4$.

\section{Comparison with the Observations}
Let us compare now the obtained results  with the observational datasets coming from CMB anisotropies. The plots contain the information we need.

In Fig. \ref{fig5}, we present the $n_{s} - r$ constraints coming from the marginalized joint 68\% and 95\% CL regions of the Planck 2018 survey and in combination with BK15 or BK15+BAO data on the non-minimal constant-roll inflationary model with a power-law coupling term. The plots are drawn for different negative and positive values of $\beta$ in the cases $N=50$ (dashed line) and $N=60$ (solid line). The panel (a) belongs to the model with a power-law coupling for the simplest form of the Hubble parameter (\ref{15}). As we can see, the model (\ref{24}) gives us large $r$ and $n_{s}$ for all considered values of $\beta$ and it seems that the model is inconsistent with the observations. By using the generalized form of the Hubble parameters (\ref{39}), we deal with the Gegenbauer polynomials so that the results are shown in the panels (a), (b) and (c) for the cases
 \begin{figure*}[!hbtp]
	\centering
	\captionsetup{width=\linewidth}
	\includegraphics[width=.48\textwidth,keepaspectratio]{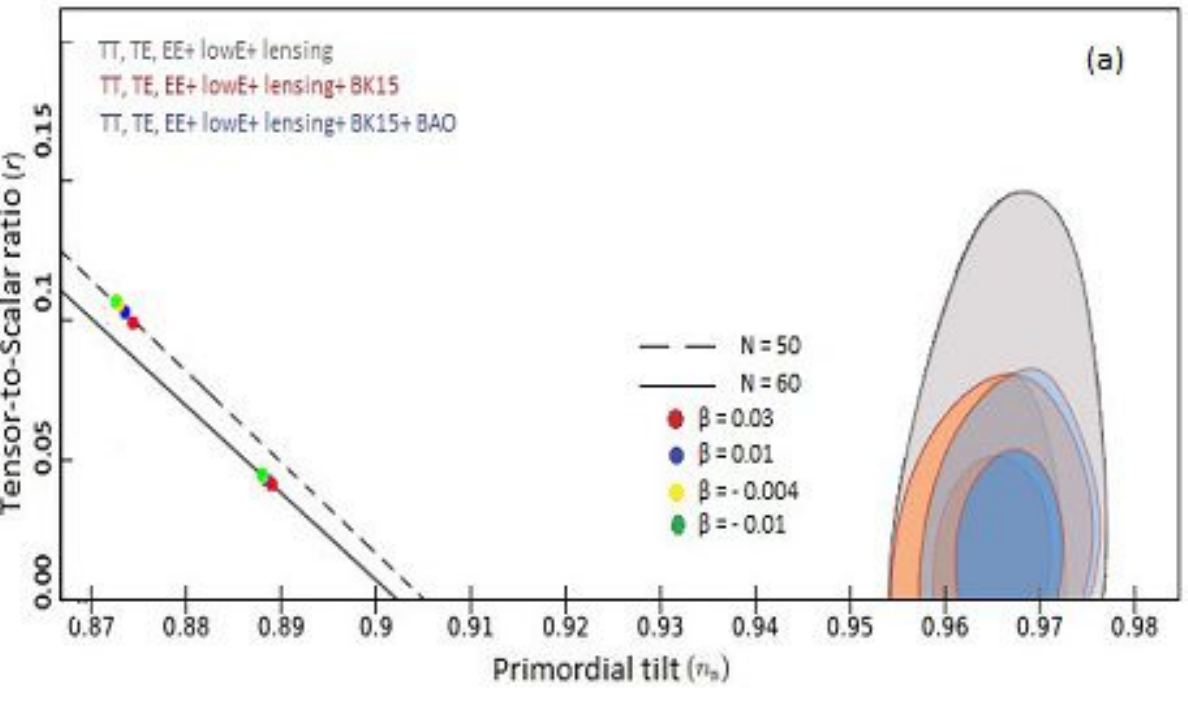}
	\includegraphics[width=.48\textwidth,keepaspectratio]{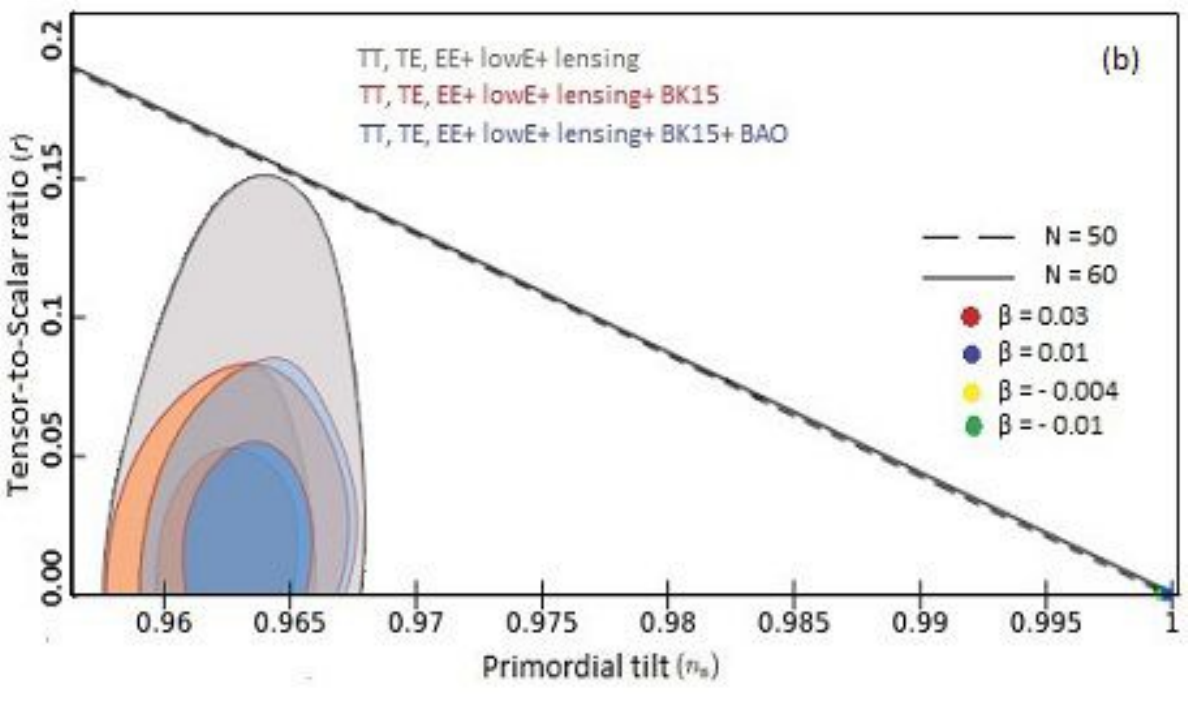}
	\centering
	\captionsetup{width=\linewidth}
	\includegraphics[width=.48\textwidth,keepaspectratio]{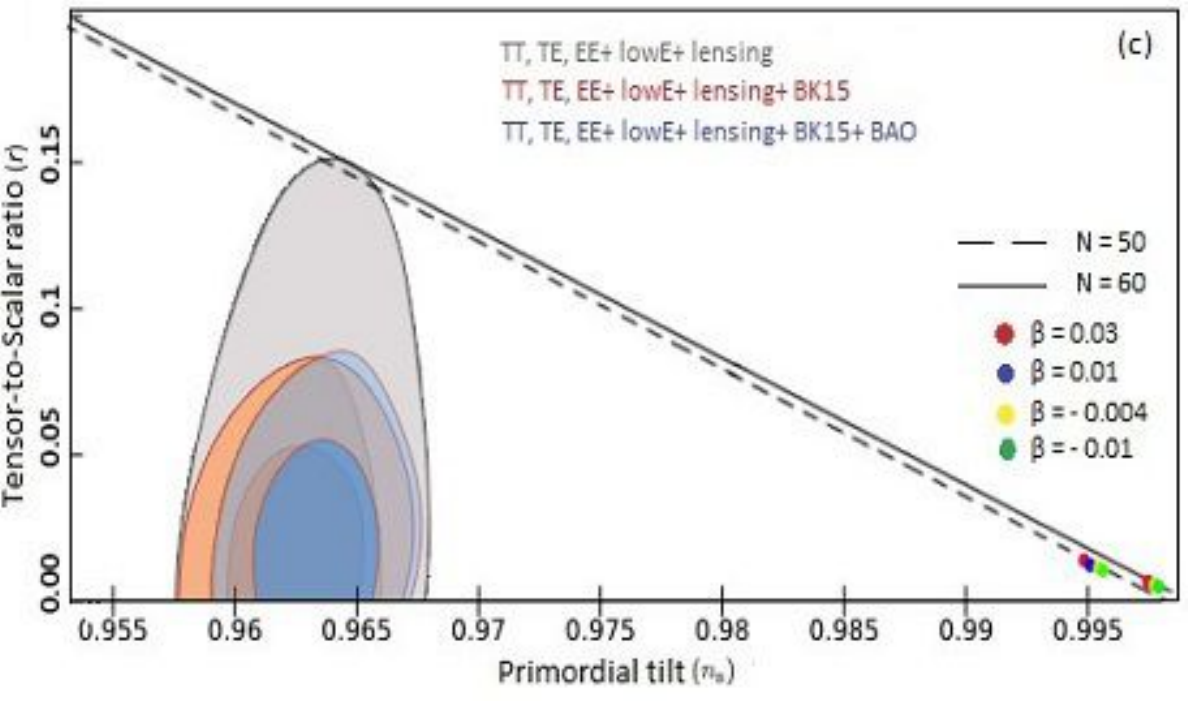}
	\includegraphics[width=.48\textwidth,keepaspectratio]{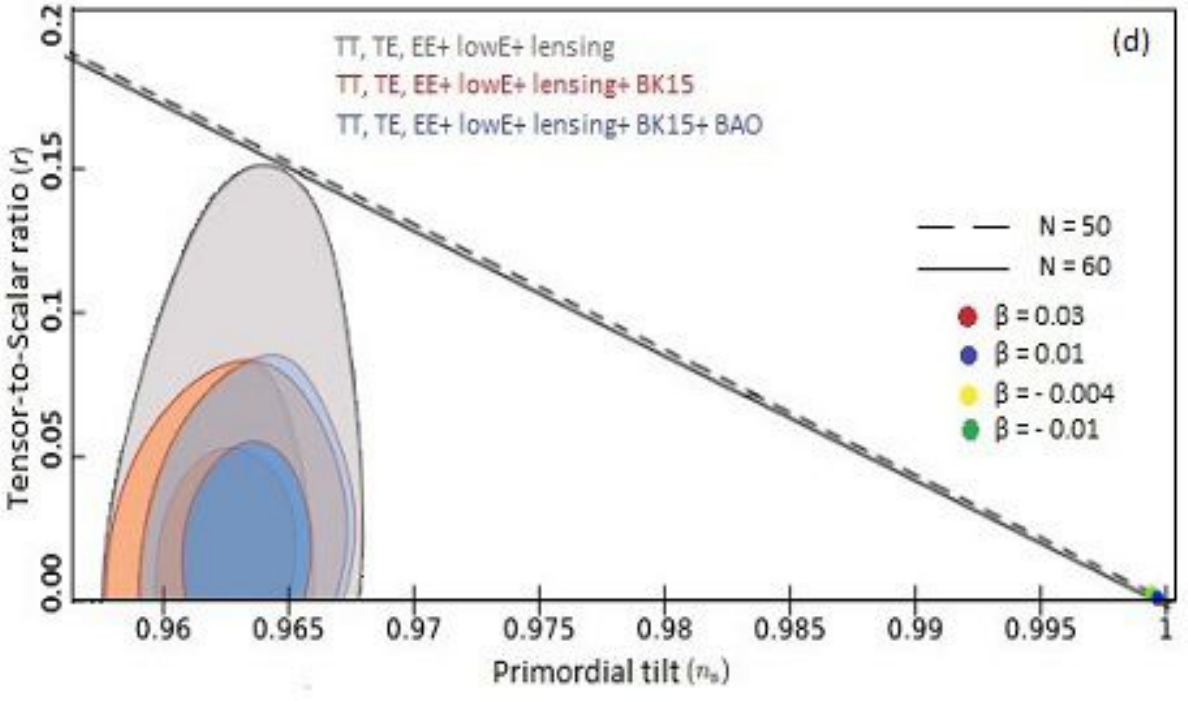}
    \caption{The marginalized joint 68\% and 95\% CL regions for $n_{s}$ and $r$ at $k = 0.002$ Mpc$^{-1}$ from Planck alone and in combination with BK15 or BK15+BAO data \cite{cmb} and the $n_{s}-r$ constraints on the model with an exponential coupling (\ref{51}). The dashed and solid lines represent $N=50$ and $N=60$, respectively. The panel (a) denotes the obtained results of the exponential coupling model with the assumption (\ref{15}) for negative and positive values of $\beta$ with $k=-0.36$. The panels (b), (c) and (d) denote the obtained results of first three polynomials $J^{\alpha}_{0}$, $J^{\alpha}_{1}$ and $J^{\alpha}_{2}$ respectively for the exponential model with the assumption (\ref{62}) for negative and positive values of $\beta$ with $\gamma=1$, $\alpha=-4$ and $k=-0.36$.}
	\label{fig6}
\end{figure*}
$P^{\lambda}_{0,2}$, $P^{\lambda}_{1,2}$ and $P^{\lambda}_{2,2}$, respectively. From the panel (c), we find that the model (\ref{45}) presents a much better prediction $r$ and also $n_{s}$, in particular, for $N=50$. The panels (b) and (c) show an inconsistency with the observations because of the large values of $r$ for the different values of $\beta$.

In Fig. (\ref{6}), we show the $n_{s} - r$ constraints form Planck 2018 and its combinations with BK15 and BAO on the non-minimal constant-roll inflation with an exponential coupling. Similar to the power-law case, we study the model for negative and positive values of $\beta$ in two cases $N=50$ and $N=60$. In panel (a), we find the constraints on $n_{s}$ and $r$ of the model (\ref{53}) with the simplest assumption of the Hubble parameter (\ref{15}). As we can see the obtained values of $r$ in the case of $N=60$ is a good agreement with the observations and also it shows a value around $0.90$ for the spectral index. The panels (a), (b) and (c)  show the constraints on the model (\ref{68}) for first three polynomials $J^{\alpha}_{0}$, $J^{\alpha}_{1}$ and $J^{\alpha}_{2}$, respectively. Obviously, the behaviour of three polynomials is very similar and, as we can see, they present an observationally consistent value of $r$ with a spectral index $n_{s}<1$.

\section{Discussion and Conclusions}
The realization of inflation is one of the most important issues of modern cosmology because, by such a paradigm, it is possible to solve many problems  at early epochs. It is mainly related to the possibility to avoid the Standard Cosmological Model shortcomings and to achieve a cosmic history  matching the observations, in particular those related to the Cosmic Microwave Background. 

Despite the robustness of the paradigm,  the  inflationary models are countless and none, up to now, is capable of addressing simultaneously all the issues related to the whole cosmic evolution, ranging from primordial quantum perturbations,  the duration of inflationary epoch, the presence of topological defects, up to the large scale structure formation and evolution. 

Besides the description and the solution of dynamical problems, inflationary models should be related to some fundamental theory as strings or some Quantum Gravity approach in order to be self-consistent.
 and physically motivated. 
 
 In this framework, NMC is assuming a relevant role due to the fact that any inflationary scenario has to face the issue of interactions between gravity, and then geometry, and the scalar field driving the inflation.
 
 Here, we proposed an approach to deal with the constant-roll inflation as modulated by the NMC in the first-order formalism.
 We showed that, depending on the form of the NMC, it is possible to achieve suitable scalar field potentials which can give rise to realistic inflationary models. 
 The main ingredient of such an approach is the relation between the Hubble rate $H$ and the inflaton modulated by two auxiliary functions $W(\varphi)$ and $Z(\varphi)$. This assumption allows to reduce dynamics to first-order cosmological equations and then to tune the slow roll of the inflaton.
 
 By comparing the proposed NMC models with data coming from Planck 2018, BK15 and BAO, it is evident that exponential coupling is preferred giving rise to realistic models. This feature is extremely relevant because it is possible to connect the related inflationary models to effective theories coming from fundamental theories like strings in the framework of the so-called Swampland Conjecture \cite{Micol,Claudio}.
 
 Clearly, the approach can be extended to other models like those coming from  higher-order gravity of other alternative theories.
 In a forthcoming paper, we will develop these studies.
 
\section*{Acknowledgments}
We would like to thank D. Bazeia for the useful comments on the manuscript. SC acknowledges the support of {\it Istituto Nazionale di Fisica Nucleare}, {\it iniziativa specifica} QGSKY.
\bibliographystyle{ieeetr}
\bibliography{biblo}
\end{document}